\documentclass[aps,pre,twocolumn]{revtex4-1} 
\usepackage{amssymb,amsfonts,amsmath}
\usepackage{graphicx}
\usepackage{float}   
\usepackage{epsfig}
\usepackage{epstopdf}   
\usepackage{verbatim}   
\usepackage{color}      
\usepackage{subfigure}  
\usepackage{hyperref}  
\usepackage{soul}
\usepackage{amsmath}
\usepackage{dcolumn}   
\usepackage{bm}        
\usepackage{amssymb}   
\usepackage[normalem]{ulem}
\usepackage[usenames,dvipsnames,svgnames,table]{xcolor}
\usepackage{lipsum,kantlipsum}

\begin{document}
\begin{center}
\title{Anomalous dynamics in tracer-particle motions in an electrohydrodynamically driven oil-in-oil system}
\author{Somayeh Khajehpour Tadavani}
\author{Anand Yethiraj}
\affiliation{Physics and Physical Oceanography, Memorial University of Newfoundland,\\
 St. John's, NL, A1B 3X7, Canada}

\begin{abstract} 
We characterize the super-diffusive dynamics of tracer particles in an electrohydrodynamically driven emulsion of oil droplets in an immiscible oil medium, where the amplitude and frequency of an external electric field are the control parameters. In the weakly-driven electrohydrodynamic regime, the droplets are trapped dielectrophoretically on a patterned electrode, and the driving is therefore spatially varying. We find excellent agreement with a $\langle x^2 \rangle \sim t^{1.5}$ power law, and find that this superdiffusive dynamics arises from an underlying displacement distribution that is distinctly non-Gaussian, and exponential for small displacements and short times. While these results are comparable with a random-velocity field model, the tracer particle speeds are in fact spatially varying in 2 dimensions, arising from a spatially varying electrohydrodynamic driving force. This suggests that the important ingredient for the super-diffusive $t^{1.5}$ behaviour observed is a velocity field that is isotropic in the plane and spatially correlated. Finally, we can extract, from the superdiffusive dynamics, a experimental lengthscale that corresponds to the lateral range of the hydrodynamic flows. This experimental length scale is only non-zero above a threshold ion mobility length.
\end{abstract}

\maketitle
\end{center}

\section{Introduction}
Diffusion is characterized by a mean-squared displacement that increases linearly with time: in 1 dimension, $\langle x^2 \rangle = 2 D t$, and the underlying distribution of displacements is Gaussian. However, in many systems in colloid science and biology, it has been recognized that a relation of the form $\langle x^2 \rangle \sim t^{\gamma}$ holds, with $\gamma \ne 1$. This is often termed as anomalous transport. Apart from simple ballistic motion (corresponding to $\gamma = 2$), there are two categories of observed anomalous motion: sub-diffusive ($0 < \gamma < 1$) and super-diffusive ($\gamma > 1)$.

Anomalous transport that is sub-diffusive or super-diffusive can be observed in biological cells \cite{metzler_gaussianity_2017, poschke_anomalous_2016, hofling_anomalous_2013,tolic-norrelykke_anomalous_2004}. The essential ingredients that generate anomalous motions are macromolecular crowding (which can restrict motions) and active driving (which can enhance motions). Sub-diffusive motions have been extensively studied in the context of macromolecular crowding \cite{banks_anomalous_2005, hofling_anomalous_2013, jeon_protein_2016}.

Enhanced, superdiffusive dynamics, while less common, is particularly relevant to (active or driven) systems out of equilibrium. A number of theoretical studies have found reasons for super-diffusive behaviour. Ajdari \cite{ajdari_transport_1995} considered a model where particles diffuse until they are reversibly adsorbed onto active filaments which propel them. Bouchaud {\it et al.} \cite{bouchaud_superdiffusion_1990}, Zumofen {\it et al.} \cite{zumofen_enhanced_1990} and Redner \cite{redner_superdiffusion_1990} found superdiffusive motions arises when there are random, but spatially correlated velocity fields. These motions, in the context of the random velocity model, arise from a probability distribution of the form $P(x,t) \sim t^{-3/4}e^{-(x/t^{3/4})^{\delta}}$, where $\gamma = 1.5$, but $\delta$ is not known. In a numerical study of L\'evy random walks  \cite{trotta_numerical_2015}, which describe non-diffusive transport that is characterized by a coupling between free-path length and free-path duration, Trotta {\it et al.} found superdiffusive transport with $\gamma \simeq 1.47$; the probability distribution function is a modified Gaussian for short displacements but a power law for large displacements.

Following tracer-particle motions inside an eukaryotic cell, Caspi and coworkers \cite{caspi_enhanced_2000, caspi_diffusion_2002} found enhanced diffusion, likely arising from microtubule-associated motor-driven motions rather than thermal motions, with an exponent $\gamma \sim 1.5$.
Ott {\it et al.}~\cite{ott_anomalous_1990} also found enhanced diffusion in a system of polymerlike micelles, where the enhancement could be killed by decreasing the breaking time of the micelle by increasing temperature. Gal {\it et al.} observe enhanced motions in polymer particles that are imbibed into living cancer cells \cite{gal_experimental_2010}, while Reverey {\it et al.} \cite{reverey_superdiffusion_2015} see superdiffusion in intra-cellular motion in highly crowded cytoplasm.

Transitions from sub- to super-diffusive are seen in systems where the forcing exceeds a threshold that overcomes pinning on a surface \cite{khoury_weak_2011}. Various groups have also reported on Brownian diffusion that can co-exist with a non-Gaussian probability distribution of displacements \cite{chechkin_brownian_2017, chubynsky_diffusing_2014, guan_even_2014, kwon_dynamics_2014, wang_anomalous_2009, wang_when_2012}, which possibly arise from competing effects at short and long times. Quite generally, this area of research is very active, and there is a call \cite{metzler_gaussianity_2017} for well-defined model systems that exhibit anomalous motions {\it and, at the same time} allow single-particle tracking of tracer-particle dynamics.

The focus of this study is tracer-particle motions in an oil-in-oil system driven by electrohydrodynamic driving forces. It has been shown previously \cite{varshney_multiscale_2016, khajehpour_tadavani_tunable_2017} that one sees both steady circulations as well as unsteady motions as a function of the driving electric field, with the unsteady motions dominating at high amplitudes leading to chaotic motions and multi-scale (seemingly turbulent) flows. In this work, we trap the oil droplets on a dielectrophoretic lattice, generating a spatially varying but periodic driving force. Given the high degree of control, the length scales on the tens of micrometers, and the presence of soft liquid-liquid interfaces, this system might be a clean and well-characterizable analog for out-of-equilibrium biological systems.

\section{Methods and techniques}
\subsection{Sample preparation and hardware} 
\begin{figure}[t] 
\centering 
\includegraphics[scale=0.19]{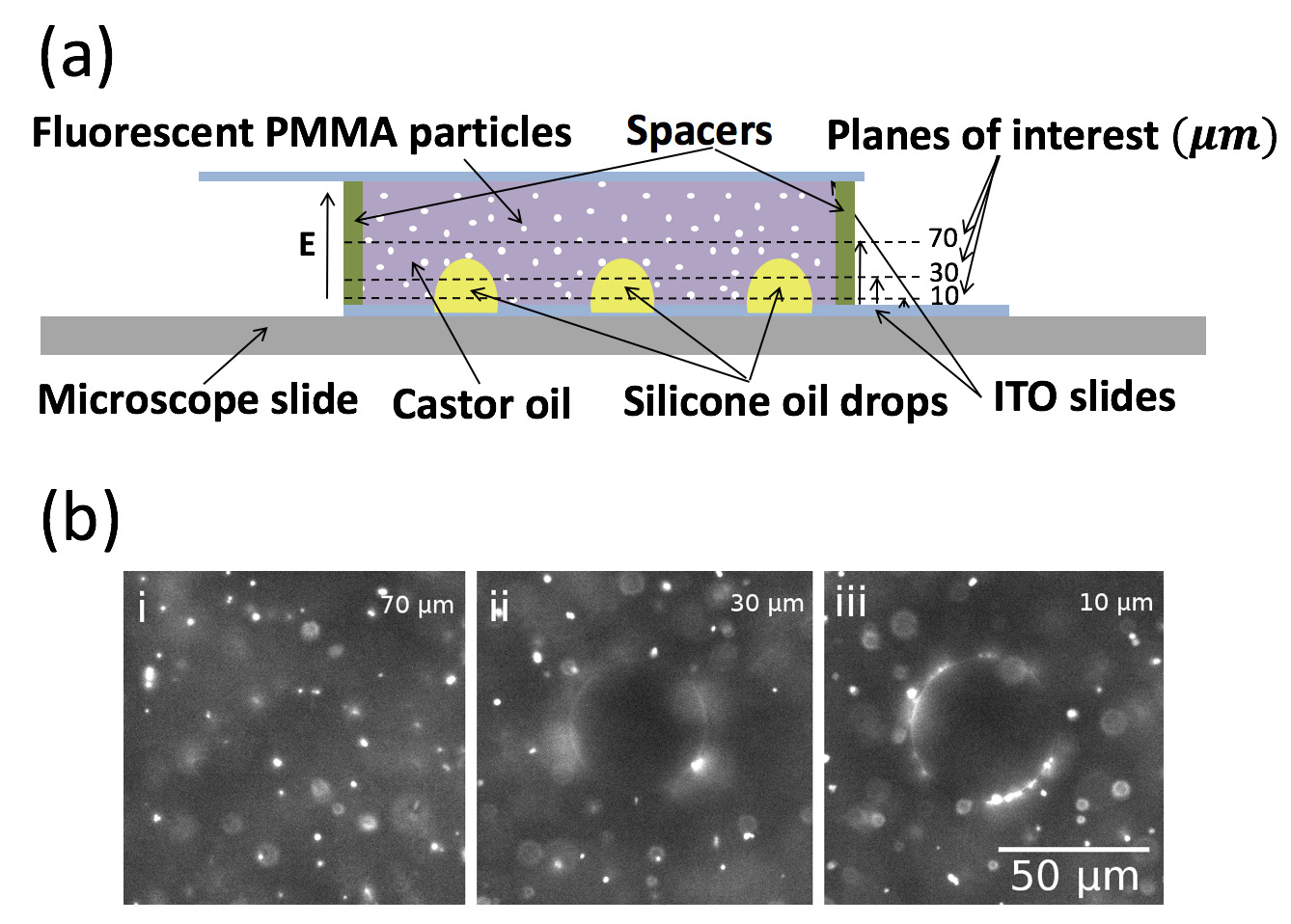}
\caption{(a) A Side view of the cell geometry. Three dash lines indicate three planes at which the experiment is conducted. The top plate is an unpatterned ITO electrode. The bottom electrode is patterned with circular  ITO-free regions. Silicone oil is dielectrophoretically trapped in the ITO-free regions and assumed the form of near-hemispherical drops (shown in yellow). The continuous phase is castor oil, seeded with fluorescent PMMA particles (represented by white dots). (b) Snapshots of three planes of interest at (i) $h_{p} = 70 $, (ii) $30$, and (iii) $10 \;\mathrm{\mu m}$ above the patterned (bottom) electrode at $E=4.6 \;\mathrm{V/\mu m}$, perpendicular to the page, and $f=1\;\mathrm{Hz}$. The PMMA particles are the bright spots in each frame. Accumulation of PMMA particles at the interface of silicone oil drop and castor oil leads to a bright edge.}
\label{FIG:Methods} 
\end{figure}

FIG.~\ref{FIG:Methods}(a) shows the side view of a sample cell with the electric field parallel to the page and to gravity. Two cover glass slides coated with indium tin oxide, ITO, are separated by glass spacers. The distance between the ITO slides is $h= 140 \; \mathrm{\mu m}$. The bottom electrode is selectively etched out, using maskless patterning and photo-lithography, in the form of a hexagonal array of 
 circular ITO-free regions, such that the diameter of each circle is approximately $ 50 \; \mathrm{\mu m}$ and the nearest-neighbor center-to-center spacing, $d_{nn}$, is approximately $100 \; \mathrm{ \mu m}$ \cite{khajehpour_tadavani_tunable_2017}. Ultraviolet-curable epoxy, Norland Optical Adhesive $61$ and $68$ are used to hold all parts together. 

The cell is filled by pipetting an emulsion of silicone oil, dielectric constant $\varepsilon_{in}/\varepsilon_{0} = 2.4 $, conductivity $ \sigma_{in} = 3.95 \times 10^{-11}\;\mbox{S/m}$, and viscosity $\mu_{in} = 0.137 \; \mbox{Pa $\cdot$ s}$, and castor oil, dielectric constant $\varepsilon_{ex}/\varepsilon_0 = 3.6 $, conductivity $ \sigma_{ex} = 4.0 \times 10^ {-10}\; \mbox{ S/m} $, and viscosity $\mu_{ex} = 0.819\; \mbox{ Pa $\cdot$ s}$, in a volume ratio of $1:16$. The subscripts $in$ and $ex$ are used to represent the (silicone oil) droplet and the (castor oil) suspending carrier fluid, respectively. Fluorescent PMMA particles with diameter of $1\;\mathrm{\mu m}$ are added to the carrier fluid, castor oil, in order to track the fluid flow. 

The silicone oil drops are prepared, prior to recording the experiment, so that they are trapped at the ITO-free regions due to the negative dielectrophoretic force \cite{khajehpour_tadavani_tunable_2017}. The experiment is carried out at four field amplitudes $3.5$, $4.6$, $5.8$, and $7.0\;\mathrm{V/\mu m}$ and for different sets of frequencies. These fields and frequencies are chosen with reference to the field-frequency phase diagram discovered in previous work \cite{khajehpour_tadavani_tunable_2017}, and all correspond to a region of parameter space that was identified as the ``ordered regime'' where the electrohydrodynamic flows are not strong enough to disrupt the hexagonal order of the trapped drops. Given that fluid flows in a leaky dielectric are governed by the boundary conditions at the interfaces, controlling the location of the interfaces enables us to follow tracer particle flows in the carrier fluid in a well-defined geometry. The experiment at each frequency and field amplitude is done at three different heights of sample cell, $10$, $30$, and $70 \mu m$ above the patterned electrode, as shown by planes of interest in FIG.~\ref{FIG:Methods}(a). FIG.~\ref{FIG:Methods}(b) shows a silicone oil drop surrounded by PMMA particles in castor oil and at three heights of interest.

A sinusoidal AC voltage, Tektronix model $AFG3022$  is amplified by a high voltage amplifier, Trek model $PZD2000A$, whose output is applied the sample cell. Fluorescence microscopy is used to observe and record the image sequences.

\subsection{Image processing}
For each experiment, a stack of 400 images, each with an exposure time of $4 \;\mathrm{ms}$, was recorded with a water-cooled digital sCMOS camera (pco.edge 5.5) and an inverted microscope, Nikon Eclipse TE2000-U. The centroid, $\vec{r'} = \left( x', y' \right) $, of each PMMA particle is first identified in each image. 

The particle trajectory in time is then obtained by standard particle tracking methods described by Crocker and Grier \cite{crocker_methods_1996, crocker_particle_nodate} using code programmed in $IDL$. Using the particle tracking information one can obtain the mean-squared displacement in 2 dimensions, $MSD =  \langle r^2(t) \rangle \equiv \langle (r'(t) - r'(t_{0}))^2 \rangle$. $\vec{r}(t) = \left( x(t) , y(t) \right)$ (where $x(t) \equiv x'(t') - x'(t'_0)$ and $y(t) \equiv y'(t') - y'(t'_0)$ respectively) is the in-plane displacement of each particle as a function of time $t = t' - t'_0$, from a variable reference start time $t'_0$.  In the above, $\left\langle \cdot \right\rangle$ is an average over all the particles in the system, and also over the reference time. We are unable to access out-of-plane motions in real time, but we obtain the same 2D information at 3 depths in the sample.

The histogram of displacements, $P(x, t)$ of all particles 
is calculated for a stack of $400$ frames and for all field amplitudes, frequencies, and heights. The same was done for $P(y, t)$.

\begin{figure*}[t] 
\centering 
\includegraphics[scale=0.2]{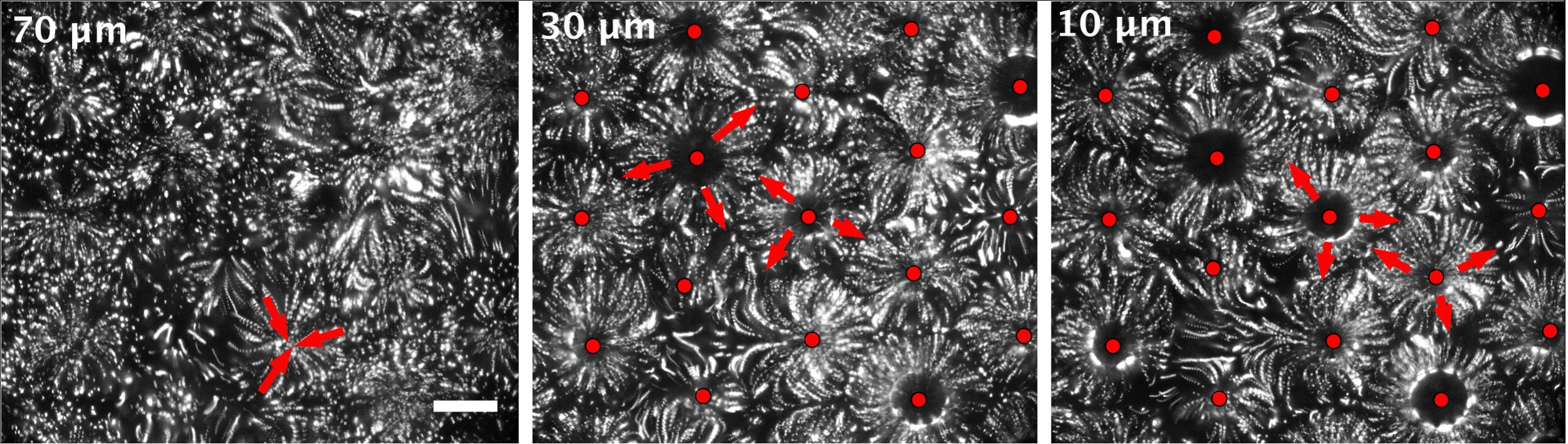}
\caption{A timelapse series of $5\; \mathrm{s}$ for three heights of interest, from left to right $h_{p}= 70$,  $30$, and $10 \;\mathrm{\mu m}$. Flows tracked by fluorescent PMMA particles are indicated by trails in direction of arrows from sources to sinks. For $h_{p}= 70\;\mathrm{\mu m}$ sources are at $d_{nn}/2 \simeq 50 \;\mathrm{\mu m}$ and sinks are at center of the drops. For $h_{p}= 30$ and $10\;\mathrm{\mu m}$ sources and sinks are located at edge of the drops and $d_{nn}/2 \simeq 50 \;\mathrm{\mu m}$, respectively, indication an oblate drop deformation. Red circles show position of center of drops in the lower heights.  $E= 5.8\;\mathrm{V/\mu m}$, perpendicular to the page, $f=3\;\mathrm{Hz}$, and scalebar is $50\;\mathrm{\mu m}$. }
\label{FIG:Prep} 
\end{figure*}

\subsection{Pattern formation and flow visualization}
Prior to recording the experiments, the system (i.e., the droplet array) was prepared as follows. The electric field was set to $E = 7.0\;\mathrm{V/\mu m}$ and $f=0.05\;\mathrm{Hz}$. At very low frequency, in the regime of strong hydrodynamics  where the hydrodynamic length, $l_{h}$, is on the order of hundreds of micrometers \cite{khajehpour_tadavani_tunable_2017}, large silicone oil drops are broken into many tiny droplets vigorously. The spontaneous breakup of droplets is a result of overcoming the viscous  stresses by electric stresses  at the interface of the droplets \cite{taylor_disintegration_1964, khajehpour_tadavani_effect_2016,varshney_multiscale_2016,varshney_self_2012}. The strong inhomogeneous flows, with breakup accompanied by coalescence, can be achieved by either lowering the frequency or increasing the field amplitude. At a fixed field amplitude, the frequency is increased to $f= 3\;\mathrm{Hz}$.  Increasing the frequency decreases the strength of the electrohydrodynamic flows, but does not change significantly the dipolar contribution. The silicone oil drops coalesce, and are attracted to the nearest ITO-free regions by negative dielectrophoresis and an array of hexagonal silicone oil droplets, in castor oil, is created on the top of patterned ITO slide \cite{varshney_large_2014}. 

All experiments are carried out at frequencies where the drops are always trapped on ITO-free regions. In steady state, toroidal flows known as Taylor vortices \cite{taylor_studies_1966, melcher_electrohydrodynamics_1969} are generated inside and outside of each droplet. In the context of the leaky dielectric model\cite{taylor_studies_1966, melcher_electrohydrodynamics_1969}, these flows are generated by the tangential component of the electric stress
at the interface of silicone oil drops and the surrounding castor oil medium; these are flows that arise from the accumulation of free charge carriers at these liquid-liquid interfaces. The PMMA fluorescent particles were used as tracer particles in order to visualize the outside flows: studying the nature of these flows is the main focus of this work. 

FIG.~\ref{FIG:Prep} shows a timelapse of $100$ frames, equivalent to $5\;\mathrm{s}$, of PMMA particles motion at three heights of interest. At $70\;\mathrm{\mu m}$ above the patterned electrode (left panel), which is roughly $20\;\mathrm{\mu m}$ above the top of the silicone oil drops, particle trajectories in the plane are less clear than for the lower heights which are chosen to contain the silicone oil drops. Due to differences in the size of droplets, the flow pattern looks disordered and asymmetric. At lower distances from the patterned electrode, $30$ and $ 10\;\mathrm{\mu m}$, middle and right panels, respectively, the flows are clearly initiated at the interface of silicone oil  and castor oil. At steady state and for most of frequencies studied in this work, drops are oblate semi-ellipdoids with major axis perpendicular to the direction of the external electric field and the circulation patterns are generated from equator to the poles \cite{taylor_studies_1966, melcher_electrohydrodynamics_1969}, as shown in FIG.~\ref{FIG:Prep}. Supplemental Material Movie 1 shows an example of PMMA tracer-particle motions around a silicone oil drop at three planes, corresponding to three heights $h_p$, with particles at each height distinguished \emph{via} false color as red, green and blue.

\section{Background and Theory} 
The mean-squared displacement or the second moment of a Gaussian distribution, in one dimension, is defined by 
\begin{equation}
\langle (x(t))^2 \rangle = \int_{-\infty}^{\infty} x^{2}P_{G}( x,t)dx= 2Dt,
\end{equation}
where
\begin{equation}
P_{G}(x,t)=\dfrac{1}{\sqrt{2 \pi D t}}\;\exp\bigg(\dfrac{-x^{2}}{2Dt}\bigg).
\end{equation}
$P_{G}(x,t)$, $D$, and $t$ represent the Gaussian probability distribution of displacements $x$, diffusion coefficient with dimension $\left[ D \right] =\mathrm{m^{2}s^{-1}}$, and time, respectively \cite{klafter_beyond_2008, chechkin_brownian_2017}. The Gaussian-diffusive trajectories are characterized by irregular, but small and homogeneous steps.

Diffusion processes in many complex systems do not follow Gaussian statistics and the mean-squared displacement (MSD) does not vary linearly in time: this is expressed (in $d$ dimensions) by 
\begin{equation}
MSD = 2 d K_{\gamma}t^{\gamma},
\label{eq:Kgamma}
\end{equation}
where $K_{\gamma}$ is a generalized diffusion constant with dimension  $\left[K_{\gamma}\right] = \mathrm{m^{2}s^{-\gamma}}$ and $\gamma$ is the anomalous diffusion exponent, with sub-diffusive motions corresponding to $0<\gamma<1$ and super-diffusive motions corresponding to $1 < \gamma < 2$.

The MSD is simply the second moment of the underlying probability distribution of displacements. Various anomalous diffusion processes yield fractional ($\gamma \ne 1$) dynamics \cite{metzler_random_2000,metzler_restaurant_2004}. The simplest generalization to the Gaussian probability distribution function of Brownian motion is given by the models of fractional Brownian motion as well as the generalized Langevin equation, which both yield
\begin{equation}
P(x,t)=\dfrac{1}{\sqrt{4\pi K_{\gamma} t^{\gamma}}}\;\exp\bigg(\dfrac{-x^{2}}{4K_{\gamma}t^{\gamma}}\bigg),
\label{EQ:FBM}
\end{equation}
where $0<\gamma\leq2$.

Enhanced diffusion with a $3/2$ power law has been related to the random velocity field model, first described by Matheron {\it et al.} \cite{matheron_is_1980} to understand water transport in microscopically heterogeneous rocks. This model describes coupling between diffusion and random, but spatially correlated velocities: a particle diffuses in a stratified fluid where each layer of fluid has a random velocity.
Redner \cite{redner_superdiffusion_1990}, Zumofen {\it et al.} \cite{zumofen_enhanced_1990}, and Bouchaud {\it et al.} \cite{bouchaud_superdiffusion_1990} obtained the probability distribution of displacements for diffusion in random velocity fields,  
\begin{equation}
P(x,t)=\dfrac{A}{\sqrt{t^{3/2}}}\;\exp\bigg(- \bigg(\dfrac{x^{2}}{4K_{\gamma}t^{3/2}}\bigg)^{\delta/2}\bigg).
\label{EQ:RVF}
\end{equation}
In the above, $A$ is the normalization factor with dimension $[A]=  m^{-1}s^{3/4}$, and $K_{\gamma}$ is the generalized diffusion constant. $\delta$ is a parameter that is not fully determined within the model, but is expected to be less than 1.7, while the asymptotic behaviour considering large displacements suggests $\delta \le 4/3$ \cite{redner_superdiffusion_1990}.

The deviation of the distribution function from a Gaussian also can be quantified using both the second moment, $\langle x^{2}\rangle$, and the fourth moment, $\langle x^{4}\rangle$, of the distribution of displacements (in 1D) using the non-Gaussian parameter \cite{rahman_correlations_1964}
\begin{equation}
\alpha_{2} = \dfrac{\langle x^{4}\rangle}{3\langle x^{2}\rangle^{2}}-1.
\label{Eq:alpha_two}
\end{equation}
The value of $\alpha_{2}$ is a sensitive measure for the type of anomalous diffusion. For Gaussian-distributed displacements $\alpha_{2}  = 0$, while this parameter deviates from zero for non-Gaussian distributions. 

In the tracer-particle tracking experiments in this work, we can readily obtain mean-squared displacements as well the 
fourth moments as a function of time; we can therefore obtain the value of $\gamma$ as well as $\alpha_{2}$. For short times, we can also construct the entire probability distribution of displacements $P(x,t)$ with reasonable precision, but out-of-plane escape of particles at long times limits doing so for all times.

\begin{figure*}[!ht] 
\centering 
\includegraphics[scale=0.38]{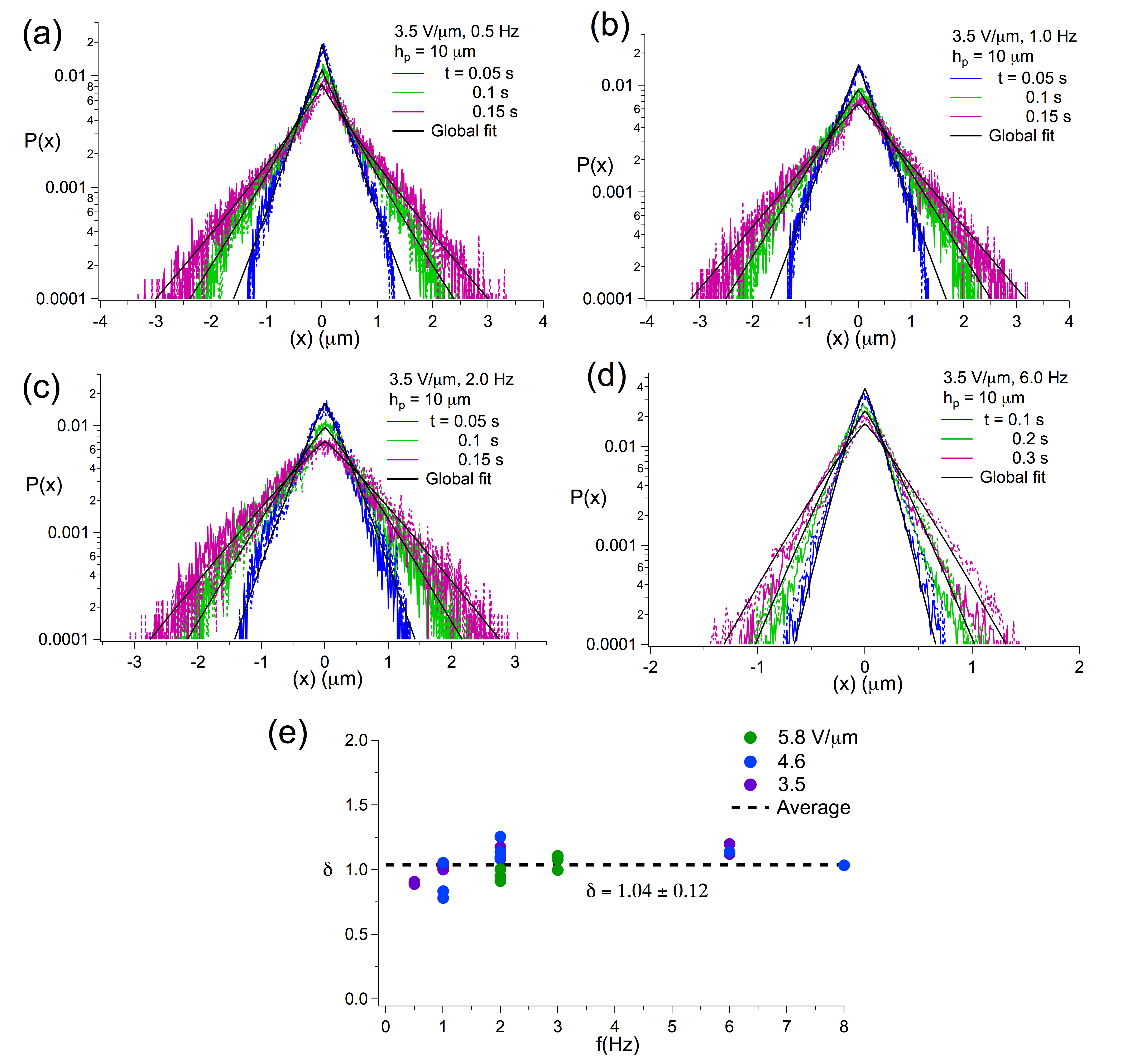}
\caption{ Normalized histogram: (a)-(d) show four examples of the normalized distribution histograms along $x$ and $y$ in three successive times and at four different frequencies, $0.5$,  $1$, $2$, and $6\;\mathrm{Hz}$, respectively. $E= 3.5\;\mathrm{V/\mu m}$, perpendicular to the page and $ h_{p}= 10 \;\mathrm{\mu m}$. (e) Variation of $\delta$ versus $f$ gives rise to mean value of $1.04\pm 0.12$.}
\label{FIG:Histograms}
\end{figure*}

\section{Results and Discussion}
\subsection{Non-Gaussian distributions}
The probability distribution of particle displacements in the $x$ and $y$ directions, as a function of $t$, is calculated based on the trajectories by averaging over all start times $t_{0}$ and particles. An example of  $ P(x,t)$ is shown in FIG. \ref{FIG:Histograms} for $E= 3.5 \;\mathrm{V/\mu m}$, at $h_{p}= 10 \;\mathrm{\mu m}$, for $0.5$, $1$, $2$, and $6\;\mathrm{Hz}$ and for different times. $P(x,t)$ only obtained for short times, as the statistics gets progressively poorer for longer times. The distribution at short times is distinctly non-Gaussian. 
Equation~\ref{EQ:RVF} is used to fit the probability distributions. For the short times examined, there is good agreement. 
Decreasing the field amplitude corresponds to lowering the driving, and so does increasing field frequency, because it takes us from the strong to the weak hydrodynamic regime (as described in earlier work~\cite{varshney_self_2012, khajehpour_tadavani_tunable_2017}.
With increasing frequency, we find that the fits are not as good at large displacements. In a few datasets, the $P(x,t)$ are asymmetric: these data were not fit. These asymmetries likely arise from drop size non-uniformities, which result in local drifts inside the cell.

From the above fits, one can get the unknown parameter $\delta$, shown in FIG. \ref{FIG:Histograms}(e). The value $\delta = 1.04\pm 0.12$ is experimentally consistent with $\delta = 1$, which is a simple exponential dependence on the absolute value of the displacement. We note, first, that these distributions have been obtained only for short times, which correspond to small overall displacements, and the behaviour at long times could be very different. While $\gamma = 3/2$ and $\delta \simeq 1$ is consistent with the prediction of $\gamma = 3/2$ and $\delta \leq 1.7 $ in the random velocity field model, we emphasize that the flows here are different. In the random velocity field model, a tracer particle samples random velocities in different strata by diffusing between strata, while in the current system, all the unsteady motions are driven motions. As shown in Supplemental Movie 2, the velocities in the plane are isotropic and spatially varying because they enter into the plane and emanate outward from each of the trapped drops (see FIG.~\ref{FIG:Prep}), and exit out of the plane in between the drops.

\subsection{Dynamics: Anomalous super-diffusive transport}
\begin{figure}[t] 
\centering 
\includegraphics[scale=0.29]{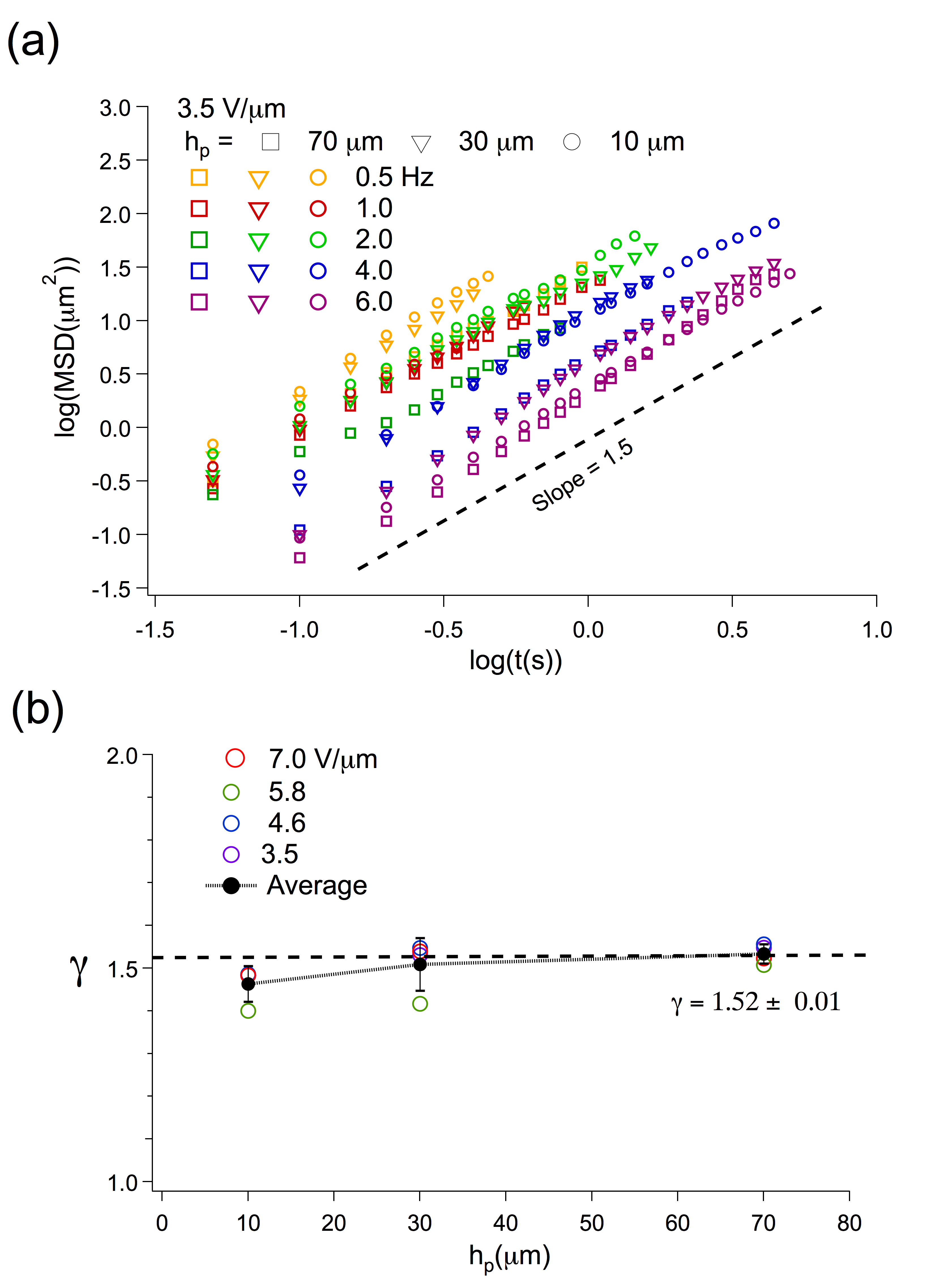}
\caption{(a) $\log(MSD)$ versus $\log(t)$ shows a linear dependency with the slope of $\gamma \simeq 1.5$ consistent with enhanced super-diffusive dynamics in the outer fluid. $\gamma \simeq 1.5$ is independent of frequency, field amplitude, and the height above the bottom electrode (i.e., the $x-y$ plane) at which the experiment is carried out. (b) $\gamma$ versus $h_{p}$, which is averaged over frequency and amplitude, respectively, shows a mean value of $\gamma = 1.52\pm 0.01$. } 
\label{FIG:MSD}
\end{figure}

Next, we examine the dynamics of particles, as a function of frequency, field amplitude and height, \emph{via} the mean-squared displacement (MSD) of tracer-particles centroids. Since the motions are isotropic in the plane (i.e., $\langle x^2 \rangle = \langle y^2 \rangle$), the 2D mean-squared displacement, $MSD = \langle x^2 \rangle +  \langle y^2 \rangle$ is plotted instead. FIG.~\ref{FIG:MSD}(a),  shows an example of $\log(MSD)$ versus $\log(t)$ for $E= 3.5 \;V/\mathrm{\mu m}$ for different frequencies and three heights of interest over a $10 \;\mathrm{s}$ time interval. 
A longer time analysis roughly about $40\;\mathrm{s}$, equivalent to a stack of $400$ frame recorded at $10 \;\mathrm{fps}$, is also carried out. Two examples, presented in FIG.1 in the Supplemental Material, show no differences between the short and long time analysis, likely because the limiting time is not the duration of observation but the time the particles remain in the plane.  For all fields, heights and frequencies, the dependence, viewed on a log-log scale, is linear and the slope is close to 3/2. The value of $\gamma$ for each field amplitude as a function of height is reported in FIG.~\ref{FIG:MSD}(b) while the average value of $\gamma = 1.52 \pm 0.01$.

There could be multiple origins for the 3/2 power law. Fractional tracer dynamics with a power law exponent $\gamma \approx 3/2$ has been reported theoretically and experimentally. A numerical study of L\'evy random walks \cite{trotta_numerical_2015} finds superdiffusive transport with $\gamma \simeq 1.47$; the probability distribution function is a modified Gaussian for short displacements but a power law for large displacements. In experiments, Caspi {\it et al.} \cite{caspi_enhanced_2000,caspi_diffusion_2002} reported super-diffusive motions with a $t^{3/2}$ scaling at short times for the MSD of a microsphere inside a living cell. They argued that a time-dependent friction imposed by the non-Newtonian medium of the living cell is responsible for the power law scaling behavior. Regner {\it et al.} \cite{regner_anomalous_2013} showed theoretically and experimentally that interacellular transport, cytoskeletal transport along microtubules, follows fractional Brownian motion with power law scaling of $3/2$ for short and long lag times.

\begin{figure}[t] 
\centering 
\includegraphics[scale=0.34]{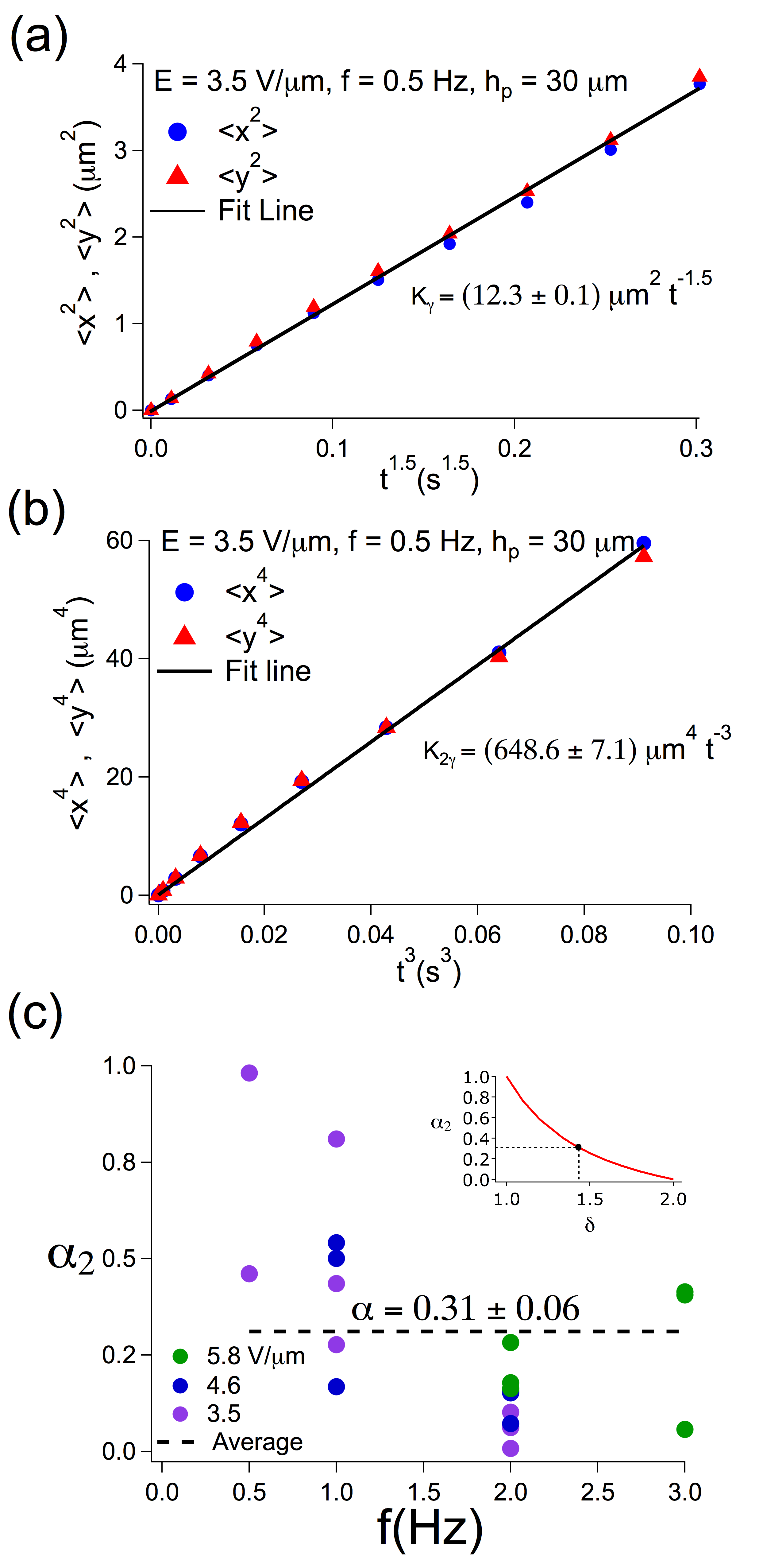}
\caption{(a) The second moment of the distribution of displacements along $x$ and $y$ depends linearly on $t^{1.5}$: the slope $K_{\gamma} = 12.3 \pm 0.1 \mu\mathrm{m^2/s^{1.5}}$ (b) The fourth moment of $x$ and $y$ distribution of displacements depends linearly on $t^{3}$: the slope $K_{2\gamma} = 648.6 \pm 7.1$. (c) The non-Gaussian parameter has no systematic frequency dependence. The average value of $\alpha_{2} = 0.31 \pm 0.06$. The inset shows $\alpha_{2}$ calculated from equation~\ref{EQ:RVF} for different values of the, in principle, unknown, $\delta$ parameter. A value $\alpha_{2} = 0.31 \pm 0.06$ corresponds to $\delta = 1.43 \pm 0.06$.}
\label{FIG:Alpha}
\end{figure}

\begin{figure*}[t] 
\centering 
\includegraphics[scale=0.29]{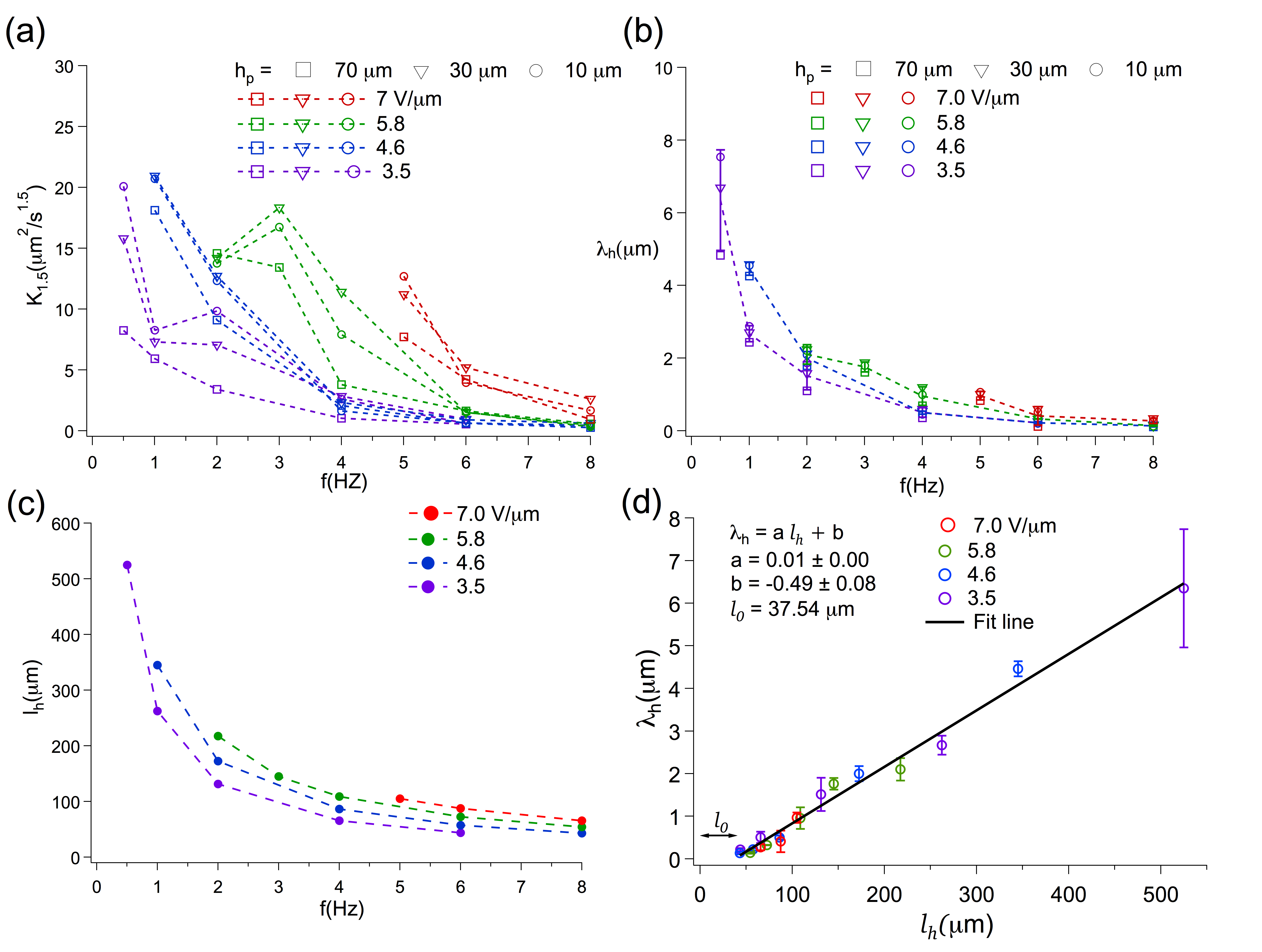}
\caption{(a) Effective diffusion coefficient, $K_{\gamma}$, versus $f$. (b) An experimental length scale, $\lambda_{h}  = \sqrt{K_{1.5}/f^{1.5}}$, versus frequency, $f$, for different field amplitudes, frequencies, and heights, shows a monotonic decrease with increasing $f$. (c) An expected ionic hydrodynamic length, $l_{h} = \mu_E E/f$, plotted for all field amplitudes for comparison with 
$\lambda_{h}$. (d) Experimental lengthscale, $\lambda_{h}$, versus ionic hydrodynamic lengthscale, $\l_{h}$, shows a linear dependency with an x intercept of $l_{0}\simeq 38 \;\mathrm{\mu m}$.} 
\label{FIG:LengthScales}
\end{figure*}

One also can measure the non-Gaussian parameter, $\alpha_{2}$, from the second and the fourth moment of the probability distribution of displacements. As a practical matter, the second and fourth moments probe large displacements more efficiently, and so this can be compared with the probability distributions obtained at short times, and hence, small displacements.
Plotted in FIG.~\ref{FIG:Alpha} (a) and (b) for one field amplitud, $E= 3.5 \;V/\mathrm{\mu m}$ and frequency $f = 0.5\;\mathrm{Hz}$, one sees that $\langle x^{2}\rangle$ and $\langle y^{2}\rangle$ have a linear relationship to $t^{1.5}$, FIG.~\ref{FIG:Alpha} (a), while $\langle x^{4}\rangle$ and $\langle y^{4}\rangle$ have a linear relation to $t^{3}$, FIG.~\ref{FIG:Alpha} (b).
Fitting to the linear forms, $\langle x^{2}\rangle = K_{\gamma} t^{1.5}$, $\langle y^{2}\rangle = K_{\gamma} t^{1.5}$ and $\langle x^{4}\rangle = K_{2\gamma} t^{3}$ and $\langle y^{4}\rangle = K_{2\gamma} t^{3}$, one can explicitly obtain, for the probability distribution in equation~\ref{EQ:RVF},
\begin{equation}
\alpha_{2} = \dfrac{K_{2\gamma}}{3K_{\gamma}^{2}}-1.
\label{Eq:New_Alpha}
\end{equation}

For all field amplitudes,  $\alpha_{2}$ is plotted in Fig.\ref{FIG:Alpha}(c) as a function of the field frequency. While the results show more dispersion than those for $\gamma$, averaging over \emph{all} datasets yields a value of $\alpha_{2} = 0.31 \pm 0.06$. For the probability distribution in equation~\ref{EQ:RVF}, $\alpha_{2}$ can be calculated given a value of $\delta$: $\alpha_{2} = 0.31 \pm 0.06$ corresponds to $\delta =1.43 \pm 0.06$.
This is larger than $\delta = 1$, obtained for short times, and interestingly, close to the asymptotic value  $\delta \simeq 4/3$ for large displacements in the random-velocity-field model~\cite{redner_superdiffusion_1990}.

\subsection{Hydrodynamic length scales}
Next, the generalized diffusion constant, $K_{\gamma}$, is shown in FIG.~\ref{FIG:LengthScales}(a). This was obtained directly from the MSD--time plot in FIG.~\ref{FIG:MSD}(a). By rewriting Eq.~\ref{eq:Kgamma} as $\log(MSD) = \log(4 K_{\gamma}) + \gamma \log(t)$, $K_{\gamma}$ was simply obtained from the y-intercept of a linear fit to each dataset in FIG.~\ref{FIG:MSD}(a).
$K_{\gamma}$ decreases with increasing frequency. This is consistent with decreasing particle diffusivity in the outer fluid. From $K_{\gamma}$, we can construct a lengthscale $\lambda_{h} = \sqrt{K_{1.5}/f^{1.5}}$. $\lambda_{h}$ is an average length one particle moves in one period of the ac oscillation, i.e., $f^{-1}\;\mathrm{s}$. $\lambda_{h}$ versus frequency is plotted in FIG.~\ref{FIG:LengthScales}(b) and shows a decreasing trend by increasing frequency.

The ion mobility length, $l_{h}$, is a characteristic length representing the range of the electrohydrodynamic interactions. $l_{h}$ can be increased by decreasing the frequency, $f$, because $l_{h} = v_{d}/f$, where $v_{d} = \mu _{E} E$ is the ionic drift velocity. $v_{d}$, in turn, is defined by electric mobility, $\mu _{E}$, and the electric field amplitude, $E$. The electric mobility is described by $\mu _{E}= z_{i}e_{0}/6\pi r\mu_{ex}$ where $z_{i}e_{0}$ is the ions’ charge, and $r$ is the radius of the ions \cite{bockris_modern_1998}. As a results, $l_{h}$ is a quantity set by  $l_{h}\propto E/f$. For castor oil, as the carrier fluid, with $\mu_{ex}= 0.819\;\mathrm{Pa.s}$, $z_{i}e_{0}= 1.6\times 10^{-19}\;\mathrm{C}$, and $r\simeq 140\times 10^{-12}\;\mathrm{m}$, the electric mobility is $\mu_{E}\simeq 75\; \mathrm{\mu m^{2} V^{-1}s^{-1}}$. FIG.~\ref{FIG:LengthScales}(c) shows $l_{h}$ versus frequency for  all field amplitudes and frequencies used in this work. $l_{h}$ at low frequencies is about $1$ order of magnitude larger than $\lambda_{h}$; however, they show a remarkably similar frequency dependence.

In FIG. \ref{FIG:LengthScales}(d), we plot $\lambda_{h}$ versus  $l_{h}$ for all field amplitudes and frequencies. $\lambda_{h}$ versus $l_{h}$ shows a remarkably linear dependence, but with an x-intercept of roughly $l_{0} \simeq 38\;\mathrm{\mu m}$. 
This indicates that $\lambda_{h}$ can be considered to be a kind of \emph{lateral} hydrodynamic length. In contrast to $l_{h}$, it is determined experimentally. $\lambda_{h}$ is only non-zero above a combination of field and frequency set by $l_{h}\propto E/f$. The $x$-intercept, $l_{0}$, is an interesting quantity that possibly indicates that there is a minimum threshold of ionic displacement required in order to initiate fluid flow. 

\section{Conclusion}
We have carefully characterized tracer motions, \emph{via} particle-tracking, in an electrohydrodynamically driven system where a spatially varying and time-dependent driving can be realized. Experiments have been carried out as a function of the amplitude and frequency of the external electric field. 

We find a robust $t^{3/2}$ power-law for superdiffusive motions in this system, and are able to extract the mathematical form of the underlying probability distribution of displacements, at least for short times. In particular, the parameter $\delta$ can be obtained both for short times, and small displacements, and for long times, and larger displacements. Its value is consistent with $\delta \simeq 1$ at short times, which corresponds to a probability distribution function with a simple exponential dependence on displacement, For longer times, we find $\delta \simeq 1.4$, close to the asymptotic value  $\delta \le 4/3$ obtained for large displacements in the random-velocity-field model~\cite{redner_superdiffusion_1990}. In this model, random velocities in different strata are coupled in by particle diffusion between strata, while in the current system, all the unsteady motions are driven motions. It is interesting that a variety of experimental systems, with different underlying driving forces, display the $3/2$-power superdiffusive behaviour. This suggests a generic origin that is insensitive to the model details. 

Finally, from the tracer motions, we find that we can extract a (frequency- and amplitude-dependent) hydrodynamic lengthscale. This lengthscale reports on the length over which tracer-particle motions are correlated, and in that sense is like the lateral (in-plane) range of hydrodynamic interactions in the system. We also measured the minimum ion mobility length, $l_{0}$, required to initiate persistent flows.

\section{Acknowledgments}
This research was supported by the National Science and Engineering Research Council of Canada (NSERC).

\newpage \newpage

\end{document}